\documentclass[usenatbib]{mnras}
\usepackage{epsfig,booktabs,caption,times,graphicx,amsmath}

\def\Hy@FixNotFirstPage{%
	\gdef\Hy@FixNotFirstPage{%
		\setbox\AtBeginShipoutBox=\hbox{%
			\copy\AtBeginShipoutBox
		}%
	}%
}
\AtBeginShipout{\Hy@FixNotFirstPage}
 
\def\I{\,\textsc{i}}
\def\II{\,\textsc{ii}}

\def\hst{{\it HST}}

\title[Pre-Explosion Counterpart of SN~2017\MakeLowercase{ein}]{A Potential Progenitor for the Type I\MakeLowercase{c} Supernova 2017\MakeLowercase{ein}}

\author[Kilpatrick et al.]{Charles D. Kilpatrick$^1$\thanks{Email:
    cdkilpat@ucsc.edu}, Tyler Takaro$^1$, Ryan J. Foley$^1$, Camille N. Leibler$^1$,
    \newauthor Yen-Chen Pan$^1$, Randall D. Campbell$^2$, Wynn V. Jacobson-Galan$^1$, 
    \newauthor Hilton A. Lewis$^2$, James E. Lyke$^2$, Claire E. Max$^1$, Sophia A. Medallon$^1$, 
    \newauthor Armin Rest$^{3,4}$\\
    $^1$Department of Astronomy and Astrophysics, University of California, Santa Cruz, CA 95064, USA\\
    $^2$W. M. Keck Observatory, 65-1120 Mamalahoa Hwy., Kamuela, HI 96743, USA\\
    $^3$Space Telescope Science Institute, 3700 San Martin Drive, Baltimore, MD 21218, USA\\
    $^4$Department of Physics and Astronomy, Johns Hopkins University, 3400 North Charles Street, Baltimore, MD 21218, USA}

\begin{document}
\date{Accepted 0000, Received 0000, in original form 0000}
\pagerange{\pageref{firstpage}--\pageref{lastpage}} \pubyear{2018}
\maketitle
\label{firstpage}

\begin{abstract}
\noindent

We report the first detection of a credible progenitor system for a Type Ic supernova (SN~Ic), SN~2017ein.  We present spectra and photometry of the SN, finding it to be similar to carbon-rich, low-luminosity SNe Ic.  Using a post-explosion Keck adaptive optics image, we precisely determine the position of SN 2017ein in pre-explosion \hst\ images, finding a single source coincident with the SN position.  This source is marginally extended, and is consistent with being a stellar cluster.  However, under the assumption that the emission of this source is dominated by a single point source, we perform point-spread function photometry, and correcting for line-of-sight reddening, we find it to have $M_{\rm F555W} = -7.5\pm0.2$~mag and $m_{\rm F555W}-m_{\rm F814W}$=$-0.67\pm0.14$~mag.  This source is bluer than the main sequence and brighter than almost all Wolf-Rayet stars, however it is similar to some WC+O- and B-star binary systems. Under the assumption that the source is dominated by a single star, we find that it had an initial mass of $55\substack{+20\\-15}~M_{\odot}$.  We also examined binary star models to look for systems that match the overall photometry of the pre-explosion source and found that the best-fitting model is a $80$+$48~M_{\odot}$ close binary system in which the $80~M_{\odot}$ star is stripped and explodes as a lower mass star. Late-time photometry after the SN has faded will be necessary to cleanly separate the progenitor star emission from the additional coincident emission.

\end{abstract}


\section{INTRODUCTION}\label{sec:introduction}

In the past three decades, there have been over $20$ detections of pre-explosion counterparts to core-collapse supernovae \citep[SNe; for a review, see][]{smartt09}.  Most of these counterparts are red supergiant (RSG) progenitor stars of Type II-P SNe (SNe with a ``plateau'' in their light curves consistent with recombination emission from an extended hydrogen envelope), which agrees with predictions from star formation and stellar evolution that suggest low-mass RSG progenitors stars should be relatively common.

There are, however, mixed results in finding the progenitor systems of other SN sub-types, with identified progenitor systems for some SNe~IIn \citep[SNe with narrow lines of hydrogen in their spectra, e.g., SNe~2005gl and 2009ip;][]{gal-yam+07,smith+10} and SNe~IIb \citep[SNe with transient hydrogen lines in their spectra, with progenitor star detections for SNe~1993J, 2008ax, 2011dh, 2013df, and 2016gkg;][]{aldering+94,woosley+94,crockett+08,maund+11,vandyk+14,dessart+11,dessart+15,kilpatrick+17}.  The progenitor stars of SNe~Ib/c (which have no hydrogen in their spectra, or helium in the case of SNe~Ic) have been comparatively elusive and only one credible pre-explosion counterpart has been identified so far in the literature \citep[the SN~Ib iPTF13bvn;][]{cao+13}.

In part, the paucity of pre-explosion SN counterparts for SNe~Ib/c is because they only make up $\sim$20\% of transients discovered in volume-limited surveys \citep[e.g., LOSS;][]{li+00,li+11,smith+11,shivvers+17}, and the incidence of SNe with deep, high-resolution pre-explosion imaging is even smaller. However, as more nearby SNe are discovered, especially those with pre-explosion {\it Hubble Space Telescope} (\hst) imaging, the growing sample of upper limits on SN~Ic progenitor systems in particular has placed strong constraints on predictions from stellar evolution and SN explosion models \citep[with deep limits on counterparts for SNe~2002ap, 2004gt, and 2007gr;][]{gal-yam+05,crockett+07,crockett+08,maund+16}. This evidence suggests that some non-RSG SN progenitor stars are either intrinsically less luminous than RSGs or heavily obscured by dust in the \hst\ optical bands typically available for pre-explosion imaging.  These stars may highly-stripped by stellar winds, and although they may be comparable in luminosity to RSGs, their SEDs peak predominantly in the ultraviolet \citep[and outside of optical or infrared bands in which pre-explosion imaging is typically available; for a review of SN~Ib/c progenitor studies see][]{eldridge+13}.  Dust obscuration is a distinct possibility for high-mass SN progenitor stars, as some high-mass RSGs are observed to have optically thick circumstellar dust \citep[e.g., SN~2012aw;][]{kochanek+12}.  In addition, high-mass SN progenitor stars ought to explode promptly, perhaps close to the dusty environments where they form \citep[see, e.g., analysis of SN environments in][]{kuncarayakti+13,galbany+16,galbany+17}.

Because SNe~Ic are the explosions of massive stars without significant hydrogen or helium in their outer layers, their progenitor star must be significantly stripped by stellar winds or a companion star. Highly-stripped Wolf-Rayet (WR) stars are therefore good candidates for SN~Ic progenitor stars \citep{yoon+10,yoon+12,yoon+17}. WR stars undergo radiatively-driven mass loss at rates exceeding $10^{5}~M_{\odot}~\text{yr}^{-1}$ \citep[although exact mass-loss rates are highly uncertain;][]{maeder+87,hamann+95,smith+14}, and so observational and theoretical evidence suggest that some pre-SN WR stars ought to be hydrogen- and helium-deficient \citep[][]{podsiadlowski+02,woosley+93,steiner+05}.

However, radiatively-driven winds are highly metallicity-dependent and WR stars tend to form in high-metallicity environments; indeed, the Small Magellanic Cloud exhibits a decreased WR-to-O-star ratio relative to Solar neightborhood \citep{hainich+15}. Predicted mass-loss rates for WR stars at Solar metallicities indicate that single WR stars may have high pre-SN masses \citep{meynet+05} and very few of these stars are predicted to be helium-poor \citep[e.g.,][]{yoon+15}.  This finding is in tension with predictions that they are SN~Ic progenitor stars given SN~Ic rates and their typical ejecta masses \citep{drout+11,taddia+15}.  One alternative is that, if WR stars are a likely channel for producing SNe~Ic, most SN progenitor systems are interacting binaries in which a WR star has been stripped by a companion.  This hypothesis is supported by the fact that many late-type WR stars are observed to be in close binaries with O-type stars \citep[e.g., WR~104;][]{tuthill+99} as well as the fact that the overall binary fraction for Milky Way WR stars is 40\% \citep{vanderhucht+01}.  If SNe~Ic mostly come from low-mass WR stars in close binaries or in dusty environments, this would explain their non-detection in optical pre-explosion imaging to date, and so examples with deep detection limits can be used to verify or rule out this possibility.

In this paper we discuss the SN~Ic 2017ein discovered in NGC~3938 on 2017 May 25 by \citet{arbour+17}.  Deep imaging starting 2~days before discovery and continuing for 2~weeks afterward indicated that SN~2017ein rose quickly after discovery \citep{atel10481}, which suggests that it was discovered very soon after explosion.

Here we report photometry, spectroscopy, and high-resolution adaptive optics imaging of SN~2017ein. We demonstrate that SN~2017ein is most consistent with carbon-rich SNe~Ic, although the source exhibits strong Na\I\ D lines at the redshift of NGC~3938 and is significantly reddened.  We use relative astrometry between our high-resolution and pre-explosion imaging, we find a single, luminous, blue source consistent with being the progenitor system of SN~2017ein, although that source appears extended and may be a blend of multiple point sources.  By comparing this source to Galactic supergiants and evolutionary tracks, we investigate channels that could produce the SN~2017ein progenitor system.

While we were preparing this manuscript, \citet{vandyk+18} published another analysis of SN~2017ein and its pre-explosion imaging.  The authors came to similar conclusions about the nature of SN~2017ein and its photometric and spectroscopic similarity to carbon-rich SNe~Ic.  They identified the same source in pre-explosion imaging as the potential progenitor system of SN~2017ein and concluded the SN likely had a very massive ($>45~M_{\odot}$) progenitor star.

Throughout this paper, we assume a distance to NGC~3938 of $m-M=31.17\pm0.10$ \citep{tully+09} and Milky Way extinction of $A_{V}=0.058$ \citep{schlafly+11}.

\section{OBSERVATIONS}\label{sec:observations}

\subsection{Archival Data}\label{sec:archival}

We obtained archival imaging of NGC~3938 from the \hst\ MAST Archive\footnote{\url{https://hla.stsci.edu/hla_faq.html}} from 11 December 2007 (Cycle 15, Proposal 10877, PI Li).  The \hst\ data were obtained with WFPC2 and consisted of two frames each of $F555W$ and $F814W$ totaling $2\times230$~s and $2\times350$~s, respectively.  We obtained the individual {\tt c0m} frames, which had been calibrated with the latest reference files, including corrections for bias, dark current, flat-fielding, and bad-pixel masking.  The images were combined using the \textsc{DrizzlePac}\footnote{\url{http://drizzlepac.stsci.edu/}} software package, which performs cosmic ray rejection, and final image combination using the {\tt Drizzle} algorithm.  The final drizzled images had a pixel scale of $0.10\arcsec$, which is consistent with the native pixel scale of the WF2 array where SN~2017ein landed. Using these final drizzled images as a reference, we used {\tt dolphot} on the individual WFPC2/{\tt c0m} frames with parameters optimised for the WF arrays on \hst/WFPC2.  These parameters were typically those recommended for {\tt dolphot} WFPC2 analysis\footnote{\url{http://americano.dolphinsim.com/dolphot}}, but with local measurements of the sky background and slightly larger aperture radius\footnote{i.e., ${\tt FitSky}=2$ and ${\tt img\_RAper}=5.33$}. This method is more accurate for faint sources in crowded fields and where non-uniform background emission can contaminate the PSFs for individual sources, as may be the case for SN~2017ein. Our combined \hst\ image is shown in \autoref{fig:astrometry}.

\subsection{Adaptive Optics Imaging}

We observed SN~2017ein in $H$ band in imaging mode with the OH-Suppressing Infrared Imaging Spectrograph \citep[OSIRIS;][]{larkin+06} on the Keck-I 10-m telescope in conjunction with the laser guide star adaptive optics (LGSAO) system on 6 June 2017. These data consisted of 2 individual frames each with 10 co-adds of $30~\text{s}$ for an effective exposure time of $300~\text{s}$ per frame or $600~\text{s}$ total. The individual frames were corrected for pixel-to-pixel variations using a flat-field frame that was created from observations of a uniformly illuminated flat-field screen in the same instrumental setup and filter.  We modeled and subtracted the sky background emission in each pixel by taking the median pixel value in a box centered on that pixel and with a width of 63 pixels (roughly 6\% the image size or 1.3\arcsec).  We masked the individual frames in order to remove bad pixels, cosmic rays, and other image artifacts.

Images taken with OSIRIS have known geometric distortions. Therefore, we calculated a distortion correction from observations of the globular cluster M92 observed on 2013 February 13 in $K$ band.  These data were reduced using the procedure outlined above, including corrections for flat-fielding using dome flats taken on the same night and in the same configuration.  We identified 65 stars common to the Gaia DR1 catalog\footnote{\url{http://gea.esac.esa.int/archive/}} and 20 frames of OSIRIS imaging, 15 of which were observed with a position angle (PA) = 0$^{\circ}$ and 5 of which were observed with PA = 45$^{\circ}$.  We fit a fifth-order polynomial to the differences $\Delta x$ and $\Delta y$ coordinate values resulting from a generalized fit of the OSIRIS imaging to the Gaia DR1 coordinates.  From this fit, we calculated a distortion correction for the OSIRIS imager\footnote{\url{https://ziggy.ucolick.org/ckilpatrick/osiris.html}}.  The standard deviation of the residual offsets of the common sources in the M92 frames was $\sigma_{\alpha}=0.011\arcsec$ and $\sigma_{\delta}=0.012\arcsec$. 

\begin{figure*}
	\setlength{\fboxsep}{-1pt}
	\setlength{\fboxrule}{2pt}
	\fbox{\includegraphics[width=0.32\textwidth]{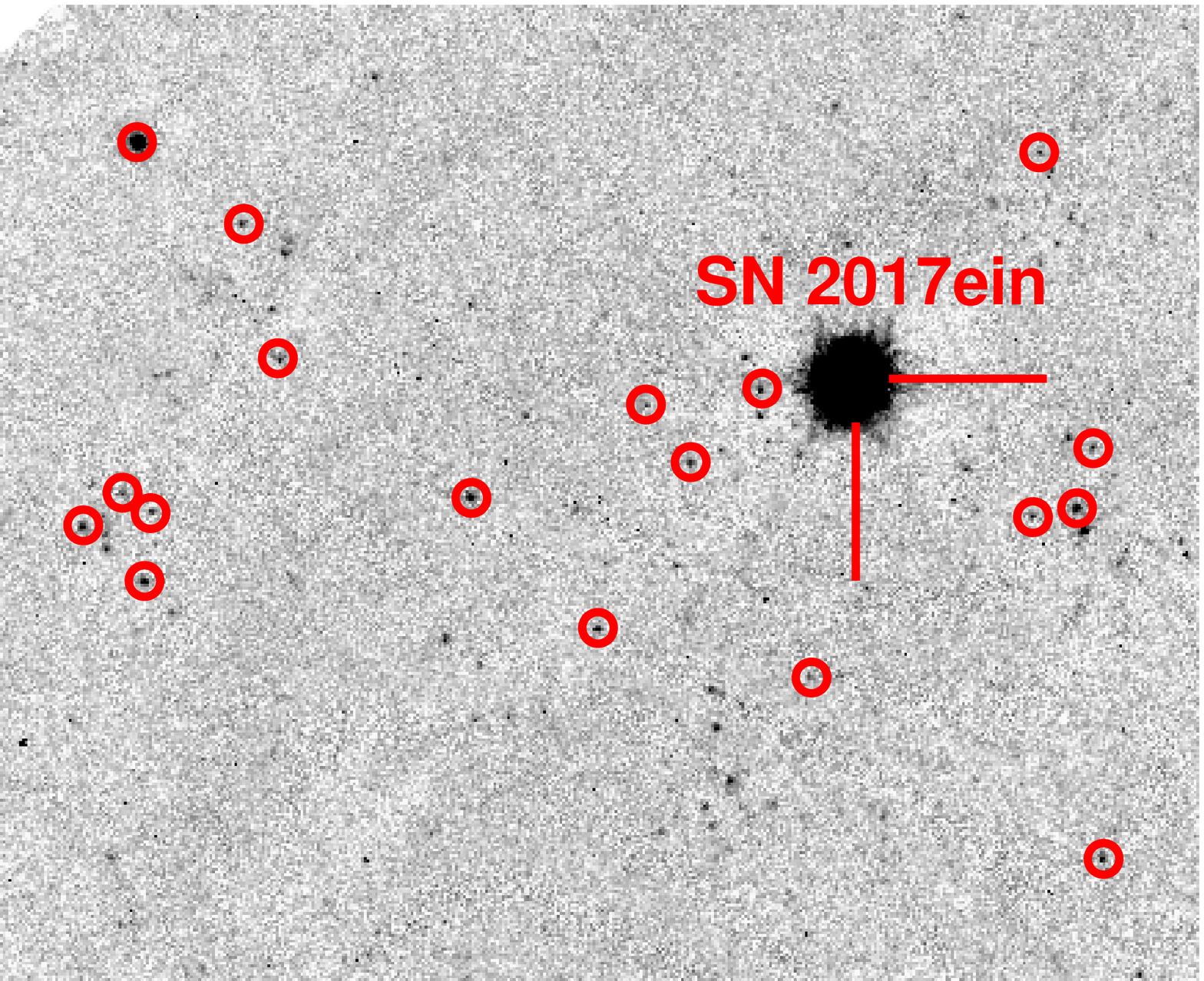}}
	\fbox{\includegraphics[width=0.32\textwidth]{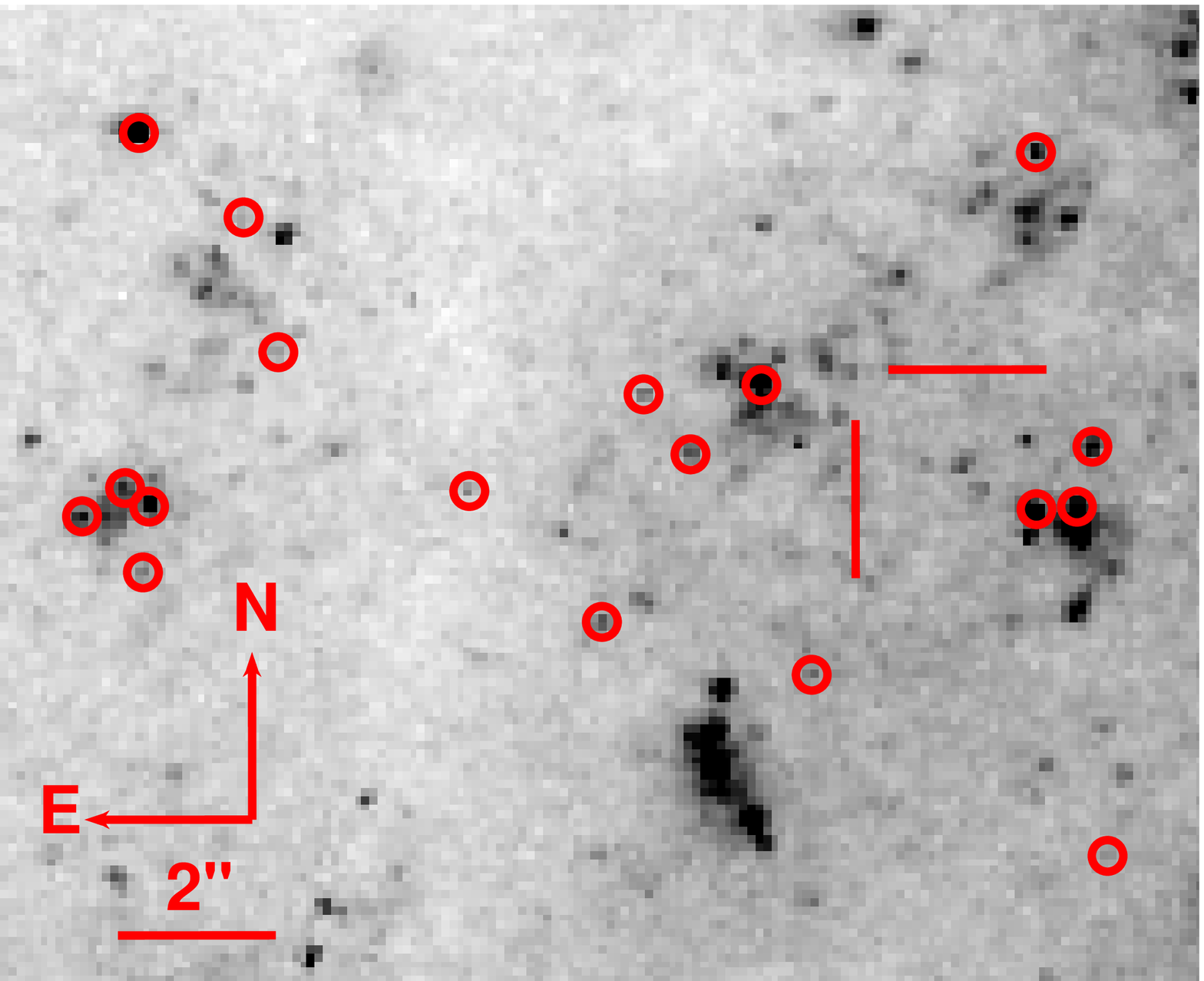}}
	\fbox{\includegraphics[width=0.32\textwidth]{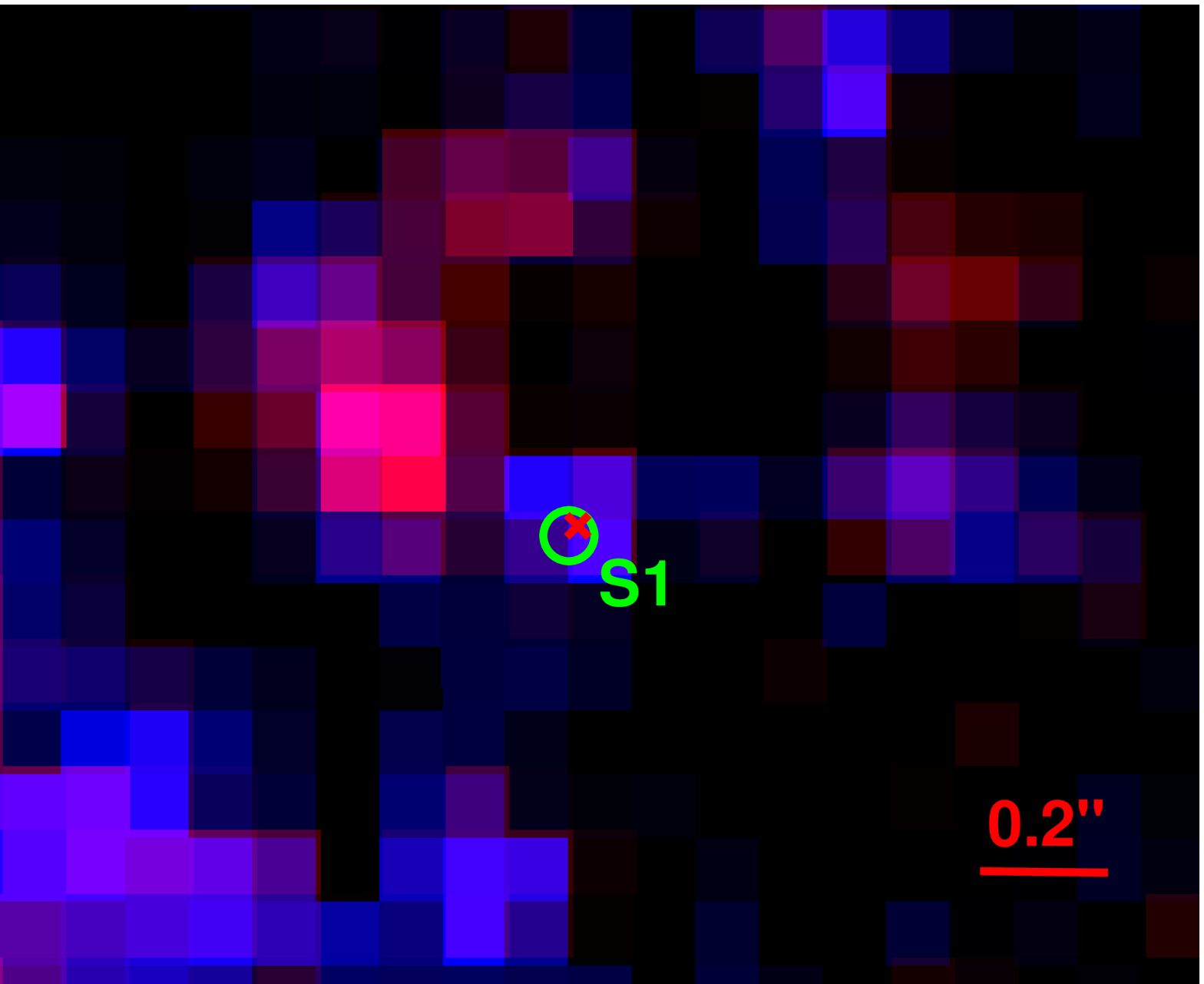}}
	\vspace{-5pt}
	\caption{({\it Left}): Keck OSIRIS LGSAO $H$-band imaging of SN~2017ein. The SN is denoted by red tick marks and 18 point sources used for astrometry are circled in red. ({\it Middle}): \hst\ WFPC2 $F814W$ reference image used for astrometry.  The progenitor star is denoted and the same 18 point sources from the OSIRIS image are circled in red. ({\it Right}): red/blue ($F555W$/$F814W$) composite \hst/WFPC2 image zoomed in on the location of the SN~2017ein counterpart. The location of SN~2017ein derived from relative astrometry along with our astrometric error ellipse is shown in green.  The coordinates of the counterpart in the $F555W+F814W$ image are shown as a red cross.  The source is relatively blue compared to the surrounding population.}\label{fig:astrometry}
\end{figure*}

Using these geometric distortion corrections, we resampled the SN~2017ein frames to a corrected grid.  Finally, we aligned the individual frames using an offset calculated from the position of the SN and combined them. In \autoref{fig:astrometry}, we show the final OSIRIS imaging along with a reference \hst/WFPC2 $F814W$ image of NGC~3938.

\subsection{Photometry}

We observed SN~2017ein with SINISTRO $g^{\prime}r^{\prime}$ imaging on the 1-m McDonald Observatory Node on the Las Cumbres Global Telescope Network (LCOGTN) from 27 May 2017 to 11 July 2017.  These data were reduced using the Obsevatory Reduction and Acquisition Control Data Reduction pipeline \citep[ORAC-DR;][]{jenness+15} using estimates of the instrumental bias, dark current, and sky flats obtained on the same night and in the same instrumental configuration. We performed PSF photometry on the final, calibrated frames using {\tt sextractor} with a PSF constructed empirically from isolated stars in each frame.  Using instrumental magnitudes from our PSF-fit photometry, we calibrated our measurements with $gr$ measurements of standard stars from the Pan-STARRS1 (PS1) object catalog \citep{chambers+16,flewelling+16}.

In addition, we observed SN~2017ein with $BVr^{\prime}i^{\prime}$ imaging using the Direct CCD on the Nickel 1-m Telescope at Lick Observatory, California.  These data were reduced using the {\tt photpipe} image reduction and photometry package.  {\tt photpipe} is a well-tested and robust pipeline used in several large-scale, optical surveys \citep[e.g., PS1 and SuperMACHO;][]{rest+05,rest+14}.  We used {\tt photpipe} to perform bias-subtraction and flat-fielding then registered the individual images.  Finally, we constructed a PSF for each image with {\tt DoPhot} \citep{schechter+93} using isolated stars in the field and then measured instrumental magnitudes for all point sources in the image.  The instrumental magnitudes were calibrated using $gri$ magnitudes from stars in the PS1 object catalog with $gri \rightarrow BV$ transformations from \citet{jester+05}.

We also obtained 8 epochs of {\it Swift} Ultraviolet/Optical Telescope (UVOT) imaging of SN~2017ein from the {\it Swift} data archive\footnote{\url{https://swift.gsfc.nasa.gov/archive/}}.  The \hst\ images demonstrate that SN~2017ein is close to several clusters and so the structure of the background near the SN is complex.  Therefore, we performed PSF photometry on the UVOT imaging using {\tt sextractor} with instrumental PSF derived from isolated stars in each UVOT image.  For images in which we detected SN~2017ein, we derived magnitudes using zero points from the most recent UVOT calibrations\footnote{\url{https://heasarc.gsfc.nasa.gov/docs/heasarc/caldb/}}.

\begin{figure}
	\includegraphics[width=0.49\textwidth]{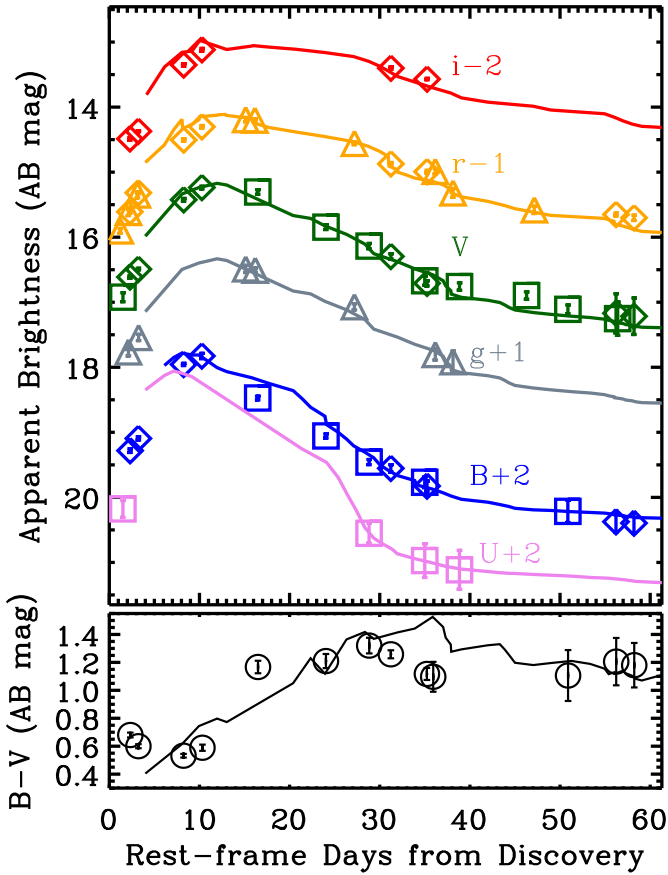}
	\caption{({\it Top}): $U\!BV\!gri$ light curves of SN~2017ein from the LCO Global Telescope Network (triangles), the Nickel telescope (diamonds), and {\it Swift} (squares).  The error bars are given inside of each plotting point.  All magnitudes are shown in the AB system with offsets between different bands.  For comparison, we overplot $U\!BV\!RI$ light curves of the carbon-rich SN~Ic 2007gr \citep{bianco+14} transformed using equations in \citet{jester+05} into the $U\!BV\!gri$ system and shifted to match SN~2017ein. ({\it Bottom}): The $B-V$ color of SN~2017ein (circles) corrected for Milky Way extinction.  We also show the $B-V$ color of SN~2007gr (black line), which has been shifted by $0.25$~mag to match SN~2017ein.}\label{fig:lc}
\end{figure}

The final photometry from all sources in $U\!BV\!gri$ is presented in \autoref{table:photometry} and in \autoref{fig:lc}.

\begin{table}
\scriptsize
\begin{center}\begin{minipage}{3.3in}
\centering {\small {\it Swift}/UVOT}\\ \vspace{5pt}
\begin{tabular}{lccc}
\hline\hline
MJD     &$U$           &$B$           &$V$ \\
\hline
57900.51 &18.177(130) &---         &16.922(069) \\
57915.46 &---         &16.472(028) &15.305(069) \\
57923.03 &---         &17.056(031) &15.846(036) \\
57927.82 &18.552(152) &17.457(039) &16.137(040) \\
57933.99 &18.972(259) &17.768(082) &16.670(050) \\
57937.84 &19.113(299) &---         &16.759(057) \\
57945.28 &---         &---         &16.897(058) \\
57949.87 &---         &18.215(172) &17.109(066) \\
57955.39 &---         &---         &17.252(256) \\
\hline
\end{tabular}
\end{minipage}
\end{center}
\begin{center}\begin{minipage}{3.3in}
\centering {\small Nickel} \\ \vspace{5pt}
\begin{tabular}{lcccc}
\hline\hline
MJD     &$B$           &$V$         &$r$          &$i$ \\
\hline
57901.30 &17.281(014) &16.613(010) &16.609(010) &16.488(011) \\
57902.22 &17.094(013) &16.491(009) &16.317(007) &16.371(010) \\
57907.25 &15.956(008) &15.425(008) &15.506(009) &15.352(009) \\
57909.26 &15.826(011) &15.237(008) &15.303(008) &15.118(009) \\
57930.23 &17.552(028) &16.295(018) &15.873(010) &15.401(010) \\
57934.24 &17.826(038) &16.709(020) &15.992(015) &15.569(013) \\
57955.20 &18.375(079) &17.169(150) &16.651(011) &---         \\
57957.20 &18.396(076) &17.216(145) &16.702(021) &---         \\
\hline  
\end{tabular}
\end{minipage}
\end{center}
\begin{center}\begin{minipage}{3.3in}
\centering {\small LCOGT} \\ \vspace{5pt}
\begin{tabular}{lcc}
\hline\hline
MJD      &$g$            &$r$  \\
\hline
57900.22  &---          &16.903(018) \\
57901.12  &16.763(030)  &16.620(024) \\
57902.12  &16.542(024)  &16.351(022) \\
57914.14  &15.488(027)  &15.192(016) \\
57915.15  &15.508(017)  &15.201(009) \\
57926.19  &16.070(024)  &15.558(010) \\
57935.16  &16.804(042)  &15.987(020) \\
57937.12  &16.915(082)  &16.319(034) \\
57946.14  &---          &16.580(028) \\
\hline
\end{tabular}
\end{minipage}
\end{center}
\caption{All $U\!BV\!gri$ magnitudes are in the AB system.  Uncertainties are given next to each measurement in milli-magnitudes.}\label{table:photometry}
\end{table}

\subsection{Spectroscopy}

\begin{figure}
	\includegraphics[width=0.49\textwidth]{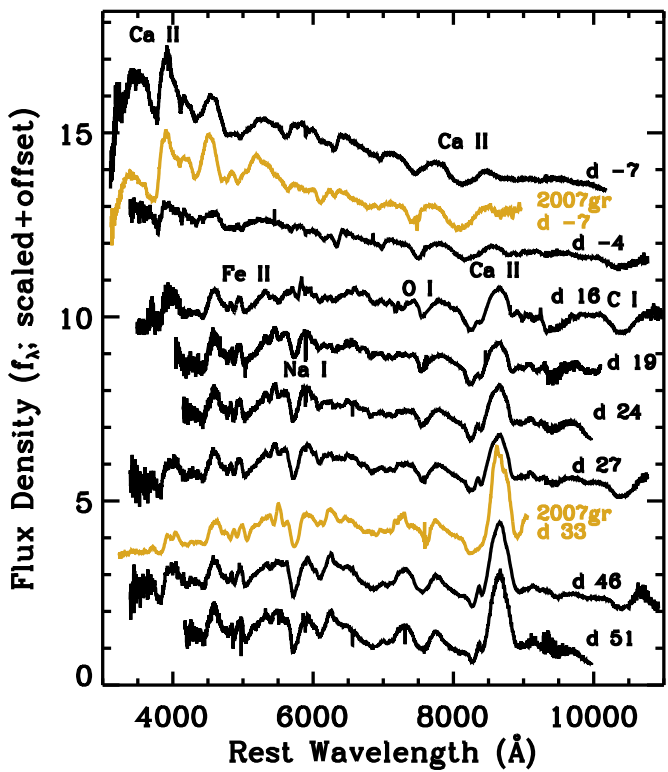}
	\caption{Our full spectral series of SN~2017ein with phase relative to $V$-band maximum indicated (as in \autoref{tab:spectra}).  The SN~2017ein spectra have been de-reddened assuming $A_{V}=1.2$~mag and using the reddening law from \citet{cardelli+89} with $R_{V}=3.1$.  For comparison, we also show two spectra of SN~2007gr from \citet{valenti+08} in the pre- and post-maximum phases.  Note the prominent carbon bands in both SN~2017ein and SN~2007gr, especially at C\II\ $\lambda$7235.}\label{fig:spectra}
\end{figure}

We observed SN~2017ein over multiple epochs (\autoref{tab:spectra}) with the Kast double spectrograph on the 3-m Shane telescope at Lick Observatory, California. The 2.0\arcsec\ slit was used and the 452/3306 grism on the blue side and 300/7500 grating on the red side in conjunction with the d57 dichroic for an approximate effective spectral range of 3400--11000~\AA\ and a spectral resolution of $R\approx400$ in each epoch. In each epoch, we aligned the slit to the parallactic angle to minimise the effects of atmospheric dispersion \citep{filippenko82}. We performed standard reductions, including bias-subtraction and flat-fielding, on the two-dimensional (2D) spectra using {\tt pyraf}. We extracted the one-dimensional (1D) spectra on the blue and red sides using the {\tt pyraf} task {\tt apall}. Wavelength calibration was performed on these 1D spectra images using calibration-lamp exposures taken in the same instrumental setup and configuration. We derived a sensitivity function and performed flux calibration using a standard star spectrum obtained on the same night and in the same setup as our SN~2017ein spectra. Finally, we combined the calibrated 1D spectra using a $\sim$100~\AA\ overlap region between the red and blue side spectra. We de-reddened each spectrum by $E(B-V)=0.018$~mag to account for Milky Way reddening and removed the recession velocity of NGC~3938 ($v = 809$~km~s$^{-1}$). The final Kast spectra are presented in \autoref{fig:spectra}.

\begin{table}
\begin{center}\begin{minipage}{3.3in}
  	\scriptsize \setlength\tabcolsep{1.0pt}
\begin{tabular}{@{}lcccccc}\hline\hline
  MJD     & Phase & Telescope/            & Range              & $R$           & Grism/        & Exposure  \\
          &       & Instrument            &                    &               & Grating       &           \\
          &                               &  &  (\AA)          &               &               & (s)       \\ \hline
57902.38  & $-$7   & Keck/LRIS            & 3120--10200        & 700           & B600/R400     & 600       \\
57906.18  & $-$4   & Shane/Kast           & 3400--10800        & 400           & B300/R452     & 1200      \\
57926.28  & 16     & Shane/Kast           & 3400--11000        & 400           & B300/R452     & 1200      \\
57929.24  & 19     & Keck/ESI             & 4051--10116        & 6000          & Echellette    & 300       \\
57934.17  & 24     & Mayall/KOSMOS        & 4160--10000        & 700           & B2K/R2K       & 900/900   \\
57937.23  & 27     & Shane/Kast           & 3400--10800        & 400           & B300/R452     & 1200      \\
57956.21  & 46     & Shane/Kast           & 3400--10959        & 400           & B300/R452     & 1200      \\
57961.17  & 51     & Mayall/KOSMOS        & 4180--10000        & 700           & B2K/R2K       & 900/900   \\
\hline
\end{tabular}
\end{minipage}
\caption{Spectroscopy of SN~2017ein. Phase is indicated in days relative to $V$-band peak brightness on MJD=57909.8.}\label{tab:spectra}
\end{center}
\end{table}

We also observed SN~2017ein with the Low Resolution Imaging Spectrograph (LRIS) on the Keck-I telescope on 2017 May 29.  We used the 600/4000 grism on the blue side and 400/8500 grating on the red side in conjunction with the D560 dichroic and the 1.0\arcsec\ long slit, which provides a spectral resolution of $R\approx700$.  SN~2017ein was observed at the parallactic angle at an airmass of $\sim$1.47.  The LRIS spectrum was reduced using standard techniques and our own IDL routines \citep[as described in][]{foley+03}.  We used a spectrum of the spectrophotometric standard HZ~44 on the blue side and BD+174708 on the red side to flux-calibrate our data and remove telluric lines from the final spectrum.  This spectra are shown in \autoref{fig:spectra}.

In addition, we observed SN~2017ein on 2017 June 30 and July 27 with KOSMOS on the KPNO 4-m telescope on Kitt Peak, Arizona.  We used the 4-pixel (1.2\arcsec) red slit in a blue setup with the B2K Volume Phase Holographic (VPH) grism (3800--6600~\AA) and a red setup with the R2K VPH grism (5800--9400~\AA).  In this setup, the spectral resolution is $R\approx700$ on both sides.  In the blue and red setups, we used the GG395 and OG530 order-blocking filters, respectively.  In both epochs, we integrated for 900~s in the blue and red setups and aligned the slit with the parallactic angle to minimise atmospheric dispersion. We performed similar reductions to the spectra described above, and the final spectra are shown \autoref{fig:spectra}.

Finally, we observed SN~2017ein with the Echellette Spectrograph and Imager (ESI) on the Keck-II telescope on 2017 June 24. We used the 0.5\arcsec\ slit with ESI and the seeing was around 0.5--0.6\arcsec\ during observations.  We observed in the echellette mode, with a spectral resolution of $R\approx6000$.  These observations were reduced using the ESIRedux IDL package \citep{prochaska+03}\footnote{\url{https://www2.keck.hawaii.edu/inst/esi/ESIRedux}}, including bias-subtraction, flat-fielding, and aperture extraction of each order using a boxcar aperture.  ESIRedux automatically applies heliocentric and barycentric corrections to the extracted spectra, and so the final spectrum is given in vacuum wavelength.  We calibrated each order using a wavelength solution derived from a ArXeHg lamp spectrum observed in the same instrumental configuration.  We calculated a sensitivity function for each order using a spectrum of BD+28~4211 observed on the same night and at a similar airmass to the SN~2017ein spectrum.  After flux calibrating each order, we combined the individual orders by taking the inverse-variance weighted average of the overlap region between the orders.  The final spectrum is shown in \autoref{fig:spectra}.

\section{RESULTS AND DISCUSSION}\label{sec:results}

\subsection{Light Curves of SN~2017ein}\label{sec:lc}

In \autoref{fig:lc}, we show our full $U\!BV\!gri$ light curves of SN~2017ein along with comparisons to $U\!BV\!RI$ light curves of SN~2007gr \citep{bianco+14} transformed using equations in \citet{jester+05} to $U\!BV\!gri$.  The light curves are all shown in the AB magnitude system and Milky Way extinction has been removed.  The comparison SN 2007gr light curves are shifted to match SN~2017ein.

\begin{figure}
	\includegraphics[width=0.49\textwidth]{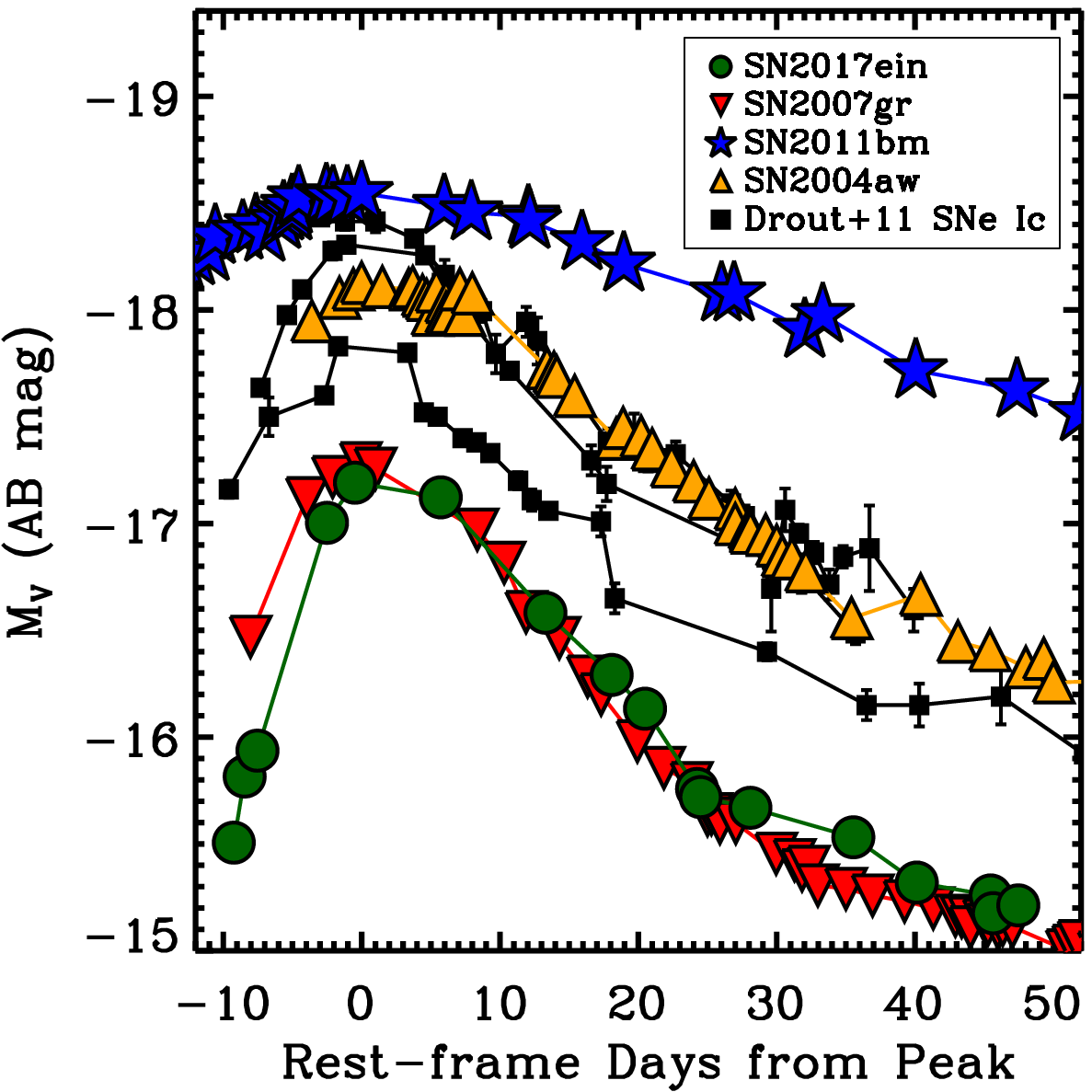}
	\caption{$V$-band light curves of SN~2017ein, SN~2007gr \citep{bianco+14}, SN~2011bm \citep{valenti+12}, and SN~2004aw \citep{taubenberger+06}.  We also plot SNe~Ic from \citet{drout+11} (SNe~2004dn, 2005fe, 2005mf) whose light curves are well-constrained near $V$-band maximum light.  The error bars on the photometry are smaller than the plotting symbols apart from the \citet{drout+11} objects, which are shown. These light curves have been de-reddened according to the values given in each reference, and with our preferred host reddening $A_{V}=1.2$~mag for SN~2017ein.  Note that the peak $V$-band absolute magnitudes of SNe~2017ein and 2007gr are at the bottom of the luminosity function for SNe~Ic, which agrees with findings from \citet{drout+11}.}\label{fig:lc-comparison}
\end{figure}

Based on our comparison to SN~2007gr, we determine that SN~2017ein peaked around ${\rm MJD} = 57909.8 \pm 1.2$ with an apparent $V$-band magnitude of $15.2\pm0.1$~mag. Removing Milky Way extinction and at the distance of NGC~3938, this value corresponds to a peak $V$-band absolute magnitude of $M_{V}=-$16.0$\pm$0.2~mag.  Without a significant level of host extinction, this luminosity implies a relatively faint SN~Ic, which typically peak from $-17$ to $-19$~mag \citep{drout+11,bianco+14}.  Even if SN~2017ein was near the bottom of the SN~Ic luminosity function derived in \citet{drout+11} and identical to SN~2007gr (e.g., with $M_{V}=-17.2\pm0.1$~mag), its host extinction would be at least $A_{V}=1.2\pm0.2$~mag. With this host extinction, SN~2017ein had a peak $V$-band luminosity dimmer than $80\%$ of SNe~Ic, which have a median $V$-band peak luminosity of $-18.0$~mag (\autoref{fig:lc-comparison}).  For example, the spectroscopically normal SN~Ic 2011bm peaked at around $M_{V}=-18.5$~mag \citep{valenti+12} whereas the SN~Ic 2004aw peaked at around $-18.1$~mag and is closer to the median for SN~Ic in terms of $V$-band magnitude \citep{taubenberger+06}.  Although there is a wide range of diversity in SNe~Ic, SN~2017ein and SN~2007gr appear to be at the bottom of the $V$-band luminosity distribution.

We also compared the host and Milky Way extinction-corrected $B-V$ color curve of SN~2007gr to the Milky Way extinction-corrected $B-V$ color curve of SN~2017ein.  The difference in $B-V$ colors is $0.25\pm0.15$~mag on average, although there is some evidence that the evolution in $B−V$ color for SN 2017ein and SN 2007gr is different. If we take this difference to be the value of $E(B-V)$ due to extinction in NGC~3938 and assume that the dust is Milky Way-like (i.e, with $R_{V}=3.1$), we find that the total host extinction is $A_{V}=0.78\pm0.47$~mag.  However, this host extinction estimate is subject to significant uncertainties, not least the assumption that the overall SN~2017ein light curve is SN~2007gr-like and the total-to-selective extinction ratio $R_{V}$.  If we assume a high value of $R_{V}=4.1$ then $A_{V}=1.0\pm0.6$~mag. In other respects, the evolution of the SN~2017ein light curves is typical for SNe~Ic.  We measure a $\Delta m_{15}$ in $r$ band of $0.6\pm0.1$~mag, which is nominally smaller but still in agreement with most other SNe~Ic including SN~2007gr \citep[e.g.,][]{drout+11,bianco+14}.

\subsection{Spectra of SN~2017ein}\label{sec:spectra}

In \autoref{fig:spectra}, we compare our spectra of SN~2017ein to spectra of the SN~Ic 2007gr \citep[in gold, from][]{valenti+08}. The comparison spectra have been de-reddened and their recession velocities have been removed according to the extinction and redshift information in \citet{valenti+08}. For each spectrum, we indicate the epoch relative to the epoch of $V$-band maximum light.  We have also de-reddened SN~2017ein by $A_{V}=1.2$~mag to account for host reddening.

\begin{figure}
	\includegraphics[width=0.49\textwidth]{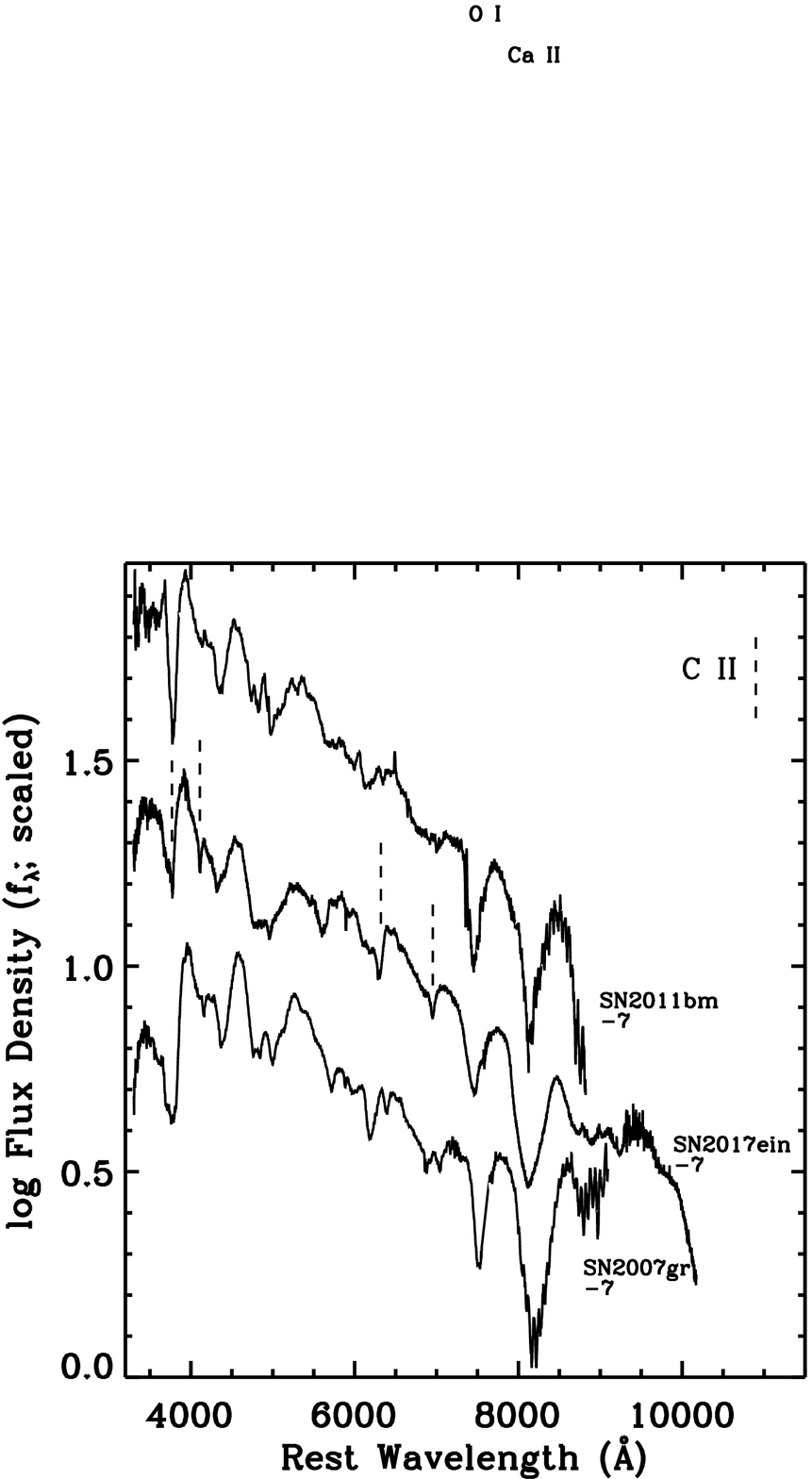}
	\caption{Comparison between our Keck/LRIS spectrum at $7$~days before $V$-band maximum to spectra of SN~2011bm \citep{valenti+12} and SN~2007gr \citep{valenti+08} from one week before optical maximum (phase is given in rest days from maximum light).  All spectra have been de-reddened and their recessional velocities have been removed according to the values given in each reference.  The SN~2017ein spectrum has been de-reddened assuming $A_{V}=1.2$~mag and using the reddening law from \citet{cardelli+89} with $R_{V}=3.1$.  We mark the presence of C\II\ features in SN~2017ein at 4267, 6580, and 7235~\AA\ as well as a possible detection at 3920~\AA.  All of these features are blueshifted by $\sim$12,000~km~s$^{-1}$.}\label{fig:comparison}
\end{figure}

At early times, it is evident that the continuum of SN~2017ein is similar to that of SN~2007gr given our choice for reddening in NGC~3938, which again indicates that SN~2017ein has a significant amount of exintction from its host galaxy.  Beyond this trend, there is clear overlap between SN~2007gr and SN~2017ein in the level of calcium and, especially, C\II\ $\lambda$7235 absorption at early times (\autoref{fig:comparison}).  These features fade over time, consistent with a decrease in temperature in the SN photosphere as it expands.  However, the similarity in the strength of this feature between SN~2007gr and SN~2017ein is noteable, especially as this carbon feature is usually weak or absent in other well-studied SNe~Ic \citep[see discussion in][]{valenti+08}.  We also note the presence of C\I\ around $10400$~\AA\ in our day 17 and 27 spectra with a weaker feature in the day $-$4 spectrum.  This evolution is consistent with a decrease in temperature in the SN~2017ein photosphere as the carbon becomes neutral. Beyond these features, SN~2017ein exhibits prominent emission lines of Fe and Ca in its post-maximum spectra.  These features are also consistent with SN~2007gr and other SNe~Ic as the SN photosphere reveals the inner layers of ejecta.

In \autoref{fig:comparison}, we compare our spectrum of SN~2017ein from $\approx$one week before $V$-band maximum to those of SN~2011bm \citep[which is a spectroscopically ``normal'' SN~Ic;][]{valenti+12} and SN~2007gr \citep{valenti+08}.  These spectra have been de-reddened and their host velocities have been removed, and we de-reddened SN~2017ein by our preferred value of $A_{V}=1.2$~mag (discussed below in \autoref{sec:extinction}) such that the continuum matches the comparison spectra. We note the similarity between SN~2007gr and SN~2017ein at this stage, especially in the presence of C\II\ absorption, which is blueshifted in SN~2017ein to a velocity of 12,000~km~s$^{-1}$.  We identify C\II\ $\lambda$4267, 6580, and 7235~\AA\ as well as a possible detection of C\II\ $\lambda$3920~\AA, although this latter feature is blended with Ca H\&K.  \citet{valenti+08} noted the presence of these features in SN~2007gr, and it is clear from \autoref{fig:comparison} that C\II\ is relatively strong in SN~2007gr at a similar epoch as SN~2017ein.  At the same time, it is clear that the SN~Ic 2011bm exhibits little or no evidence for strong carbon absorption, and so SN~2017ein may be relatively carbon-rich or else viewed at an angle such that carbon absorption was seen.

Does the presence of these carbon features reflect a high intrinsic carbon abundance or is it simply an effect of the ionisation state in the ejecta?  \citet{mazzali+10} suggest that the presence of strong carbon features in SN~2007gr reflect a high intrinsic carbon-to-oxygen ratio in the ejecta.  However, carbon was still ionised at this epoch indicating that the ejecta were still hot. Therefore, at $\approx$one week before maximum light, we are only seeing through the outermost layers of ejecta as most of the SN is still optically thick to electron scattering.  Late-time nebular spectra will be critical for evaluating the intrinsic carbon abundance in SN~2017ein and determining whether the similarity to SN~2007gr reflects an intrinsically high carbon abundance.

\subsection{Na~I D Equivalent Width and an Estimate of the Host Extinction}\label{sec:extinction}

\begin{figure}
	\includegraphics[width=0.49\textwidth]{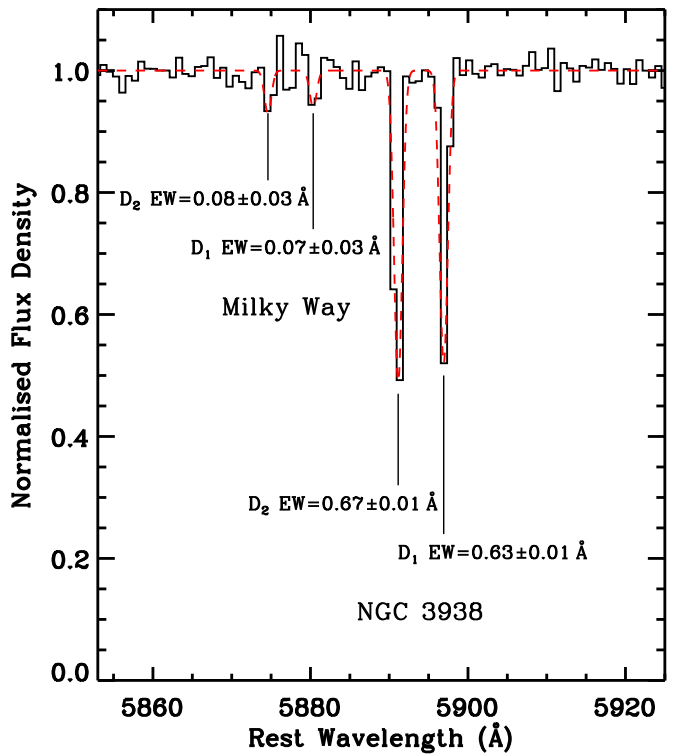}
	\caption{Our normalised Keck/ESI spectrum of SN~2017ein from 20~days after maximum light and centered around the Na\I\ D lines.  We detect Na\I\ D at a redshift consistent with both the Milky Way and NGC~3938 and measure the equivalent width in both sets of lines.  The Milky Way equivalent width is consistent with the reddening along this line-of-sight as reported in \citet{schlafly+11}.  As we discuss in \autoref{sec:spectra}, the NGC~3938 equivalent width is consistent with a very high host extinction, from $A_{V}=1.2$--$1.9$~mag depending on the total-to-selective extinction ratio.}\label{fig:naid}
\end{figure}

In addition to the unusual continuum shape, there is evidence of significant reddening in our spectra in the strong Na\I\ D features redshifted to the velocity of NGC~3938.  These lines are well-resolved in our Keck/ESI spectrum (\autoref{fig:naid}), where we estimate the equivalent width (EW) of the features at the velocity of NGC~3938 to be EW Na\I\ D$_{1}=0.63\pm0.01$~\AA\ and Na\I\ D$_{2}=0.67\pm0.01$~\AA, for a combined EW of D$_{1}+$D$_{2}=1.30\pm0.02$~\AA. Following the relation for Milky Way-like dust in \citet{poznanski+12}, we find $E(B-V)=0.47\pm0.05$~mag. We also measure the Na~\textsc{I} D EW from Milky Way lines in the ESI spectrum, which is consistent with the reddening value from \citet{schlafly+11}.

The NGC~3938 reddening value is significantly larger (at $>1\sigma$) than the value derived by comparing the SN~2017ein and SN~2007gr color curves.  It is possible that this discrepancy reflects an intrinsic difference between the SN~2017ein and SN~2007gr light curves, and so our comparison between these objects is flawed.  In this case, Na~\textsc{I} D provides a more reliable estimate of the total host extinction to SN~2017ein, although converting the value of $E(B-V)$ to an in-band extinction involves further assumptions about the host extinction properties.

Variation in dust properties in SN host galaxies is a major uncertainty, and some examples are known to have dust unlike the Milky Way \citep[e.g., SN~2014J in M82;][]{gao+15}.  For a reasonable range of total-to-selective extinction ratios (e.g., $R_{V}=2.5-4.1$), the reddening we derive from Na~\textsc{I} D could imply a host extinction to SN~2017ein from $A_{V}=1.2$--$1.9$~mag.

SN~2017ein is similar to the relatively low-luminosity SN~Ic 2007gr, and although the intrinsic $B$- and $V$-band magnitudes and color curves of these objects may be discrepant (at the $0.2$~mag level), the $V$-band host extinction to SN~2017ein implied by this comparison is $1.2\pm0.2$~mag.  This extinction value implies $R_{V}=2.6$ assuming $E(B-V)=0.47\pm0.05$~mag from Na\I\ D.  We adopt this value for the ``preferred'' extinction to the SN~2017ein progenitor system in our subsequent analysis, although there are still large uncertainties on this estimate. $A_{V}$ could be $0.7$~mag larger than this value for $R_{V}=4.1$, but this value would be inconsistent with our fits to the SN~2007gr $V$-band light curve and $B-V$ color curves, which are both consistent with lower values close to $A_{V}\approx1.2$~mag. Therefore, while we use this value throughout the rest of the paper, the total systematic uncertainty on the host extinction is large.  Reasonable estimates on the value of $A_{V}$ range from roughly $1.2$~mag (implying $R_{V}=2.5$) to $1.9$~mag ($R_{V}=4.1$).

All extinction estimates from the SN neglect the possibility of circumstellar dust that affected the progenitor observables but was destroyed by the SN. We do not see evidence for excess emission in the early-time light curve or spectroscopy of SN~2017ein that would be consistent with interaction between the SN shock and a significant mass of dust \citep[as in SNe~IIn, e.g.,][]{fox+11,kilpatrick+18}.  Considering that SN~2017ein may have been discovered very soon after explosion \citep[as suggested by][]{atel10481}, we predict that we would have observed excess emission from a large mass of dust, and so we find the presence of such dust to be unlikely.

\subsection{Relative Astrometry Between the Adaptive Optics and \hst\ Imaging}\label{sec:rel}

We performed relative astrometry between the OSIRIS image and composite \hst\ image using the 18 common sources circled in both frames (\autoref{fig:astrometry}). The positions derived for these 18 sources were determined using {\tt dolphot} in the \hst\ frame and {\tt sextractor} in the OSIRIS frame.  We performed image registration on the LGSAO image using the \textsc{IRAF} tasks {\tt ccmap} and {\tt ccsetwcs}.  We used default parameters for {\tt ccmap}, which fit pixel coordinates from the stars identified in our LGSAO imaging to a tangent plane projection of the right ascensions and declinations of the same stars in the \hst\ image.  We used a general geometric fit, which included terms for linear shift, rotation, and the relative pixel scale between the images. 

We estimated the astrometric uncertainty of our best-fitting geometric projection by randomly sampling half of the common sources and calculating a geometric solution then calculating the average offset between the remaining common sources in this projection. On average, we found $\sigma_{\alpha} = 0.040\arcsec$ and $\sigma_{\delta} = 0.037\arcsec$.  Within the combined uncertainties (totaling $\sigma_{\alpha}=0.041$\arcsec\ and $\sigma_{\delta}=0.039$\arcsec) of the relative astrometry, the position of SN~2017ein in our LGSAO image, and the geometric distortion correction, we find a single source at the location of SN~2017ein in the reference \hst\ image, which we call S1 (\autoref{fig:astrometry}). S1 is located at $\alpha= 11^{\text{h}}52^{\text{m}}53^{\text{s}}.264$, $\delta=+44^{\circ}07\arcmin26\arcsec.619$ and is detected in the combined \hst\ frame with $S/N = 46$ for an astrometric precision of $0.004\arcsec$.  This is also the same source that \citet{vandyk+18} identify as the counterpart to SN~2017ein. As we demonstrate in \autoref{fig:astrometry}, S1 is offset from the position of SN~2017ein as determined from our LGSAO image by $0.037$\arcsec, or approximately 0.75$\sigma$.  In the \hst\ image, we do not detect any other sources within a $0.343$\arcsec\ (8.6$\sigma$) radius of the position of SN~2017ein.  Therefore, we consider S1 to be the only viable candidate as the counterpart to SN~2017ein.

We estimate the probability of a chance coincidence in the \hst\ image by noting that there are a total of $102$ sources (including extended sources) with $S/N>3$ within a 10\arcsec\ radius of SN~2017ein in the \hst\ image from \autoref{fig:astrometry}.  Therefore, the fraction of the total solid angle within 10\arcsec\ of SN~2017ein that is within 3$\sigma$ of a detected source is approximately $102 \times (3 \times 0.04\arcsec / 10\arcsec)^{2} = 1.5\%$. This value represents the probability that the detected point source is a chance coincidence.  Therefore, although it is unlikely that the identified point source is a chance coincidence, there some probability that this is the case.  Follow-up imaging will be critical in order to confirm or rule out this possibility.

\subsection{Photometry and Classification of the Pre-Explosion Counterpart}

\subsubsection{\hst\ Photometry and the PSF of the Pre-Explosion Counterpart}\label{sec:psf}

From our photometric analysis of S1, we obtained Vega magnitudes $m_{F555W} = 24.787\pm0.041$~mag and $m_{F814W} = 24.902\pm0.075$~mag.  These values are nominally fainter than those found in \citet{vandyk+18}, who report $m_{F555W} = 24.56\pm0.11$~mag and $m_{F814W} = 24.58\pm0.17$~mag. These differences originate from the different {\tt dolphot} parameters used in fitting\footnote{\citet{vandyk+18} use {\tt FitSky=3} and {\tt img\_RAper=8}.}, which we adjusted to fit for the complex local background around S1.

Photometry from the combined $F555W+F814W$ imaging suggests that the object at the position of SN~2017ein has sharpness $=-0.061$ and roundness $=0.053$, which is generally consistent with a single point source. However, the PSF of the source in both $F555W$ and $F814W$ is somewhat extended and eccentric, with PSF eccentricities $0.221$ and $0.218$, respectively. The source may be partially contaminated by emission from a nearby cluster (it has {\tt dolphot} crowding parameter $0.163$; see \autoref{fig:astrometry}), hence our use of a local background estimate in performing photometry (\autoref{sec:archival}), although it is possible that this background emission somewhat affects the PSF parameter estimates.

\begin{figure}
	\includegraphics[width=0.49\textwidth]{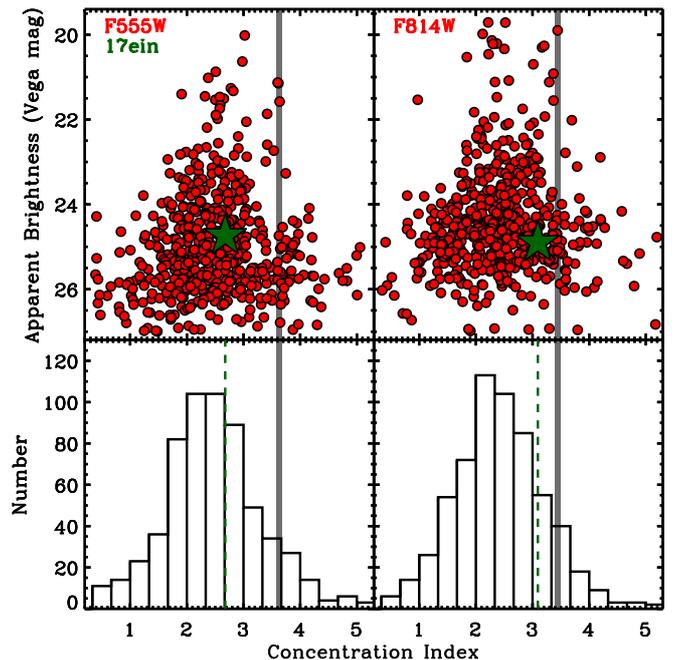}
	\caption{({\it Top left}): Concentration index versus Vega magnitude for sources in the \hst/WFPC2 $F555W$ images of NGC~3938.  The value for the S1 counterpart to SN~2017ein (\autoref{sec:rel}) is shown as a green star.  The shaded region shows the nominal split between clusters and stars at a concentration index of 3.63 in $F555W$ as discussed in \autoref{sec:psf}. ({\it Top right}): Same as the top left but for the $F814W$ image with the concentration index threshold set at 3.44.  ({\it Bottom left}): A histogram of the $F555W$ values shown in the upper left plot as a function of concentration index.  The value for S1 is shown as greek dashed lines.  ({\it Bottom right}): Same as the bottom left but for the $F814W$ values.}\label{fig:concentration}
\end{figure}

We further investigated the possibility that S1 is an extended source by measuring a ``concentration index'' (\autoref{fig:concentration}).  Following analysis in \citet{chandar+10} for \hst/WFC3 photometry, we calculated the difference in magnitudes for a circular aperture with radius $0.5$~pixels and an aperture with radius $3.0$~pixels centered on the PSF-fit coordinates and using the same local background estimate from {\tt dolphot}.  We restricted our analysis to sources within 10\arcsec\ of SN~2017ein, and so all of the sources we investigated landed on the same WF2 array as SN~2017ein. 

We found a concentration index of 2.677 in $F555W$ and 3.094 in $F814W$.  Larger concentration indices imply that most of the emission of the source is spread out at large separations from the center of the PSF.  \citet{chandar+10} define a threshold concentration index for distinguishing between stars and clusters by examining candidate objects that are thought to be stars or clusters and finding the maximum and minimum concentration indices of these distributions.  These thresholds cleanly separate sources detected using WFC3 into distributions of stars and clusters \citep[see, e.g., Figure~4 in][]{chandar+10}.  We estimate a similar threshold by examining the concentration index below which we find 95\% of objects with $m<25$~mag, which are more likely to be clusters.  This value is 3.63 for $F555W$ and 3.44 for $F814W$.  In both cases, the SN~2017ein candidate lies within the distribution and could reasonably be considered to be a single, unresolved point source.  However, in $F814W$, the source is much closer to the limit we have defined, which is relatively crude compared to the analysis in \citet{chandar+10}.

If S1 is a cluster, then it is extremely young.  The $m_{F555W} - m_{F814W}$ (roughly $V-I$) color corrected for Milky Way extinction and assuming no host extinction is $-0.143$~mag, which is consistent with a $4$~Myr cluster \citep{bruzual+03,peacock+13}.  With added host extinction, the source would be even bluer and younger, implying that any star that exploded from the population in this cluster had an initial mass $>73~M_{\odot}$ for a single star \citep[as derived from MIST evolutionary tracks;][]{choi+16}.

If S1 is a marginally unresolved cluster, then only the most extreme and massive single star populations could explain the colors for this source.  At $4$~Myr these sources would have $M_{V} < -10$~mag or an unabsorbed magnitude of $m_{F555W} \approx 21.2$~mag at the distance of NGC~3938.  Unless the host extinction is $A_{V} > 4$~mag to SN~2017ein, it is unlikely that S1 is such a star.  We find that the most likely scenario is a luminous, blue source corresponding to a single star or multiple star system with a surrounding population of less luminous sources.  These other sources are likely unresolved stars still on the main sequence.  This scenario agrees with the magnitudes, colors, and concentration for S1.

If the full {\tt dolphot} photometry for S1 is partly contaminated by a surrounding population of stars, we can remove some of this light using forced photometry.  We used the instrumental PSFs for \hst/WFPC2 in $F555W$ and $F814W$ to fit photometry to S1 and found that the central source was marginally fainter without the extended emission: the PSF-fit source, which we call PSF1 had brightnesses $m_{F555W} = 24.901\pm0.062$ and $m_{F814W}  = 25.112\pm0.121$.  If a single object dominates the emission from this source and is the pre-explosion counterpart to SN~2017ein, then these magnitudes represent its total emission.  Otherwise, if one or more unresolved sources contributes significantly to the emission within the PSF, then these magnitudes are only upper limits on the pre-explosion emission from the SN~2017ein progenitor star.

\subsubsection{Classification of the Pre-Explosion Counterpart}\label{sec:classification}

\begin{figure}
	\includegraphics[width=0.49\textwidth]{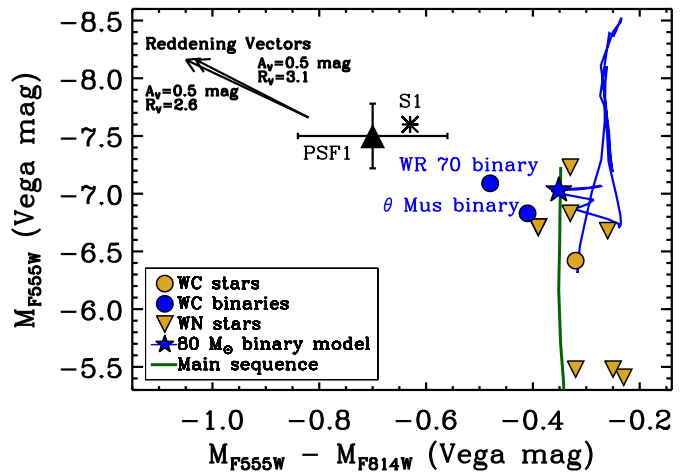}
	\caption{Color-magnitude diagram showing the properties of the forced photometry from the SN~2017ein pre-explosion counterpart (PSF1).  The values for $F555W$ luminosity, and $F555W-F814W$ color are described in \autoref{sec:classification}.  We have corrected for our preferred host reddening of $A_{V}=1.2$~mag.  The uncertainties are based on the photometry and distance modulus.  We also indicate reddening vectors corresponding to $A_{V}=0.5$~mag with $R_{V}=2.5$ and 4.1.  For comparison, we show the full photometry, including extended emission around the source S1.  We also plot a MIST model for the zero age main sequence for a population of stars at the metallicity of NGC~3938 (green line) as well as a BPASS model for a $80$+$48~M_{\odot}$ binary with initial period $P=10~{\rm days}$ (blue line).  The terminal state of the BPASS model is shown with a blue star.  Several Galactic WR systems are shown as orange triangles (WN stars) and orange circles (WC stars).  All of these WR stars represent the emission from the WR star itself apart from the WC9+B star binary WR~70 and the WC6+O binary $\theta$ Mus (blue circles; labeled), which are dominated by the spectral energy distributions of their companion stars.}\label{fig:cmd}
\end{figure}

We corrected the \hst\ photometry of PSF1 for interstellar exinction using the extinction law in \citet{cardelli+89} with $A_{V}=0.058$ and $R_{V}=3.1$ and host extinction using $A_{V}=1.2$~mag with $R_{V}=2.6$ (implying $A_{I}=0.6$~mag) as discussed in \autoref{sec:extinction}.  Assuming $m-M=31.17\pm0.10$, PSF1 had luminosities $M_{F555W}=-7.5\pm0.2$~mag (roughly $V$ band) and $M_{F814W}=-6.7\pm0.2$~mag (roughly $I$ band).  We plot these values in \autoref{fig:cmd} along with the corresponding values for S1.  Clearly the added light from extended emission does not make a significant difference to the final photometry of PSF1 relative to the uncertainties.

The extinction-corrected $F555W$-$F814W$ color is $-0.67\pm0.14$~mag (with systematic uncertainties represented by variations in reddening; \autoref{fig:cmd}). This color is blueward of the main sequence and implies a much hotter star than the vast majority of stars.  For example, the bluest $V-I$ color for a star in the Hipparcos and Tycho2 catalogue is $V-I=-0.49$, although this estimate is subject to significant selection bias due to Galactic dust.  At the very least, this color implies a spectral energy distribution that peaks far blueward of $V$ band and a source with a very hot photosphere. Some late-type WN stars \citep[also WNL; Wolf-Rayet stars with low-ionization state nitrogen lines in their spectra;][]{hamann+06,sander+12,crowther07} are luminous enough to match the properties of PSF1 with $M_{V}=-7.6$~mag, but typically only have $V-I$ colors as blue as $-0.3$~mag.

If we assume the source with $M_{F555W}=-7.5\pm0.2$~mag and $F555W-F814W$ color $-0.67\pm0.14$~mag is dominated by a single star, then by comparison to MIST models \citep{paxton+11,paxton+13,paxton+15,dotter+16,choi+16}\footnote{http://waps.cfa.harvard.edu/MIST/}, we find that the best-fitting bolometric correction for such a star is $BC_{F555W}=-2.8\pm0.4$~mag.  The best-fitting luminosity for a single star at the metallicity of NGC~3938 \citep[$\log(O/H) = 8.94\pm0.05$;][]{kudritzki+15} is approximately $\log(L/L_{\odot})=6.0\pm0.2$, implying an initial mass of $55\substack{+20\\-15}~M_{\odot}$.

The evolutionary pathway such a high-mass star would take to end as a highly-stripped SN~Ic progenitor is less certain.  It has been hypothesized that WC stars \citep[WR stars with strong carbon emission lines in their spectra;][]{crowther07} are likely candidates for SNe~Ic \citep{dessart+10,yoon+12,dessart+12,yoon15,dessart+15}.  The Galactic population of WC stars generally have absolute magnitudes $M_{V}=-3$ to $-5.5$~mag with $V-I$ colors bluer than WNL stars \citep{crowther+07,vanderhucht+01}.  Most SN~Ic progenitor star models propose that the final mass ought to be very low in order to explain the observed abundances of SNe~Ic, with $17~M_{\odot}$ at the most \citep{meynet+03,yoon15}.

We examined Galactic WR stars with detailed photometry, extinction, and distance estimates \citep[from the VIIth catalogue of Galactic WR stars;][]{vanderhucht+01} to determine whether any known stars are luminous and blue enough to match PSF1. We restricted our sample to stars without detected companions so we could examine the intrinsic colors and luminosities of WR stars.  None of the stars in our sample had colors and luminosities that matched PSF1, and only a few WN stars had luminosities or colors that approached PSF1 (examples with the most luminous $M_{V}$ magnitudes and bluest $V-I$ colors are shown in \autoref{fig:cmd}).

However, roughly 40\% of Galactic WR stars are in binaries, often where the WR star itself does not dominate the overall spectral energy distribution.  There are some examples of WR stars in binaries with O- or B-type supergiants that could agree with the luminosity and colors of PSF1.  All known examples in the VIIth catalogue of Galactic WR stars with $M_{V}<-6.5$~mag and $V-I<-0.3$~mag are late-type WC stars with O- or B-type supergiant companion stars. For example, the WC9 star WR~70 \citep[also HD~137603;][]{williams+00} is in a binary with a B0I supergiant with intrinsic $M_{V}=-7.09$~mag and $V-I=-0.48$~mag while the W6+O binary $\theta$ Mus has $M_{V}=-6.83$~mag and $V-I=-0.41$~mag \citep{moffat+77,stupar+10}.  Both of these stars could match the observed colors and luminosity of PSF1 if we decreased the amount of reddening closer to the lower limit of the range allowed by spectroscopy and photometry of SN~2017ein ($A_{V}=0.5$~mag).  

If PSF1 is dominated by light from a star other than the progenitor star of SN~2017ein, it could simply be in the same cluster as SN~2017ein.  The precision of our astrometry indicates that there could be as much as a $0.03$\arcsec\ offset between SN~2017ein and the transient source we identified in pre-explosion imaging, corresponding to $\sim$3~pc at the distance to NGC~3938.  If the star were $3$~pc away from the SN~2017ein progenitor star, it would be unassociated with the progenitor system.  On the other hand, if PSF1 is dominated by the progenitor system of SN~2017ein and the actual progenitor star is the less luminous component of a binary system, then the two components should be coeval and the luminosity and color of the more luminous source can be used to constrain the properties of the other component.

Therefore, we analyzed all Binary Population and Stellar Synthesis \citep[BPASS2.1;][]{eldridge+17} models to look for binary star models that terminate with a total luminosity $M_{V} < -7.0$~mag and color $V-I < -0.35$~mag.  We restricted our search to models with the metallicity of NGC~3938, but otherwise searched the entire range of models with primary initial mass $0.1$--$300~M_{\odot}$, initial mass ratio $0.1$--$0.9$, and initial log period (days) $0$--$4$, consisting of 12,664 models.  There were two such models in the overall sample that terminated within the selected parameter space, both with primary mass $80~M_{\odot}$.  These two models, which we will call Models 1 and 2, have initial mass ratios $0.6$ and $0.8$ and log periods $1$ and $0.8$, respectively.  Models 1 and 2 end with the secondary star comprising $\sim80\%$ of the overall $V$-band luminosity, although in both cases this star is somewhat redder ($V-I\approx-0.27$ to $-0.31$~mag) than the overall system.  The final mass of the primary star in Model 1 (wider/smaller initial mass ratio) is $15.4~M_{\odot}$ while in Model 2 (closer/larger initial mass ratio) the primary terminates with $52.2~M_{\odot}$.

If the host reddening to SN~2017ein was relatively low ($A_{V}=0.7$~mag), then the color and luminosity of PSF1 could agree with Model 1 (shown in blue in \autoref{fig:cmd}).  In addition, Model 1 terminates with effectively no hydrogen and only $0.20~M_{\odot}$ of helium.  The total helium mass fraction in the star is quite low ($<0.1$), and even for a relatively low total ejecta mass, the helium mass fraction in the ejecta would be consistent with predictions for SN~Ic progenitor stars \citep[which suggest that a mass fraction $<0.5$ is sufficient to hide helium lines][]{dessart+11,yoon+15}.

The luminosity and color of Model 1 are somewhat similar to WC+O star binaries, where the primary has undergone significant stripping and/or radiative mass loss and ends up as a relatively low mass star.  The fact that the primary star in this model has a somewhat high mass for a WC star \citep[which typically range from $4$--$9~M_{\odot}$; for a review see][]{crowther+07} could be explained by the WR mass-loss prescription or a slightly higher metallicity at the location of SN~2017ein in NGC~3938.  Alternatively, this type of binary could simply be a rare system with a high-mass WC star that explodes promptly.  Ultimately, the full implications for the initial metallicity, mass-loss prescription, and binary parameters of this system are complex, and additional modeling is needed to explore WC star binaries as potential SN~Ic progenitor stars.

\section{The Nature of SN~2017ein and Its Progenitor System}

We find a luminous, blue source (S1) at the progenitor site of SN~2017ein. The environment around S1 is consistent with the environments of SNe~Ic as a whole; these SNe preferentially explode in regions of high star formation rates \citep{galbany+14,galbany+16}, which strongly suggest a high mass ($>$25~$M_{\odot}$) progenitor star that evolves and explodes close to the region where it formed.  However, the diversity of SNe~Ic as a whole, and in particular their progenitor systems, is still poorly understood.  It is possible that SN~2017ein is atypical for SNe~Ic, implying an unusual progenitor system.

From our analysis of the pre-explosion photometry, S1 appears marginally extended, and may be consistent with a massive star cluster. Indeed, it has been found that many stripped-envelope SNe are discovered in or near such clusters \citep{fremling+16,maund+17}.  The fact that S1 is more extended in the redder $F814W$ band suggests that the surrounding population of unresolved sources come from stars with lower mass than the star or stars dominating S1. If we assume that S1 is dominated by emission from a single star, then it has a best-fitting mass of $55\substack{+20\\-15}~M_{\odot}$.  This mass range is consistent with the findings of \citet{vandyk+18}, who report the source is consistent with a star with an initial mass of $60$--$80~M_{\odot}$ at the most for binary star models, or in the range of 47--$48~M_{\odot}$ for single star models.

SN~2017ein could have exploded from a star in a multiple system where the primary does not dominate the overall spectral energy distribution.  Late-type WC stars often occur in systems with O- or B-type supergiant companion stars, such as the WC9+B binary WR~70 or the WC6+O star binary $\theta$ Mus (\autoref{fig:cmd}).  Many of these systems lack detailed orbital parameters, and so it is difficult to place strong constraints on the nature of the WC star itself.  WR~70 is a relatively high-mass WC star \cite[$9.8~M_{\odot}$]{nugis+02} and, as an late-type WC star, likely has an intrinsically high carbon abundance \citep{smith+82}.  The overall luminosity is dominated by the B-type supergiant companion star, and so the total luminosity of this particular system is a poor indicator of the mass of the WC star.  Comparison to binary star models suggests it may be possible to obtain such a system with a $80$+$48~M_{\odot}$ system, although the mass of the final component is somewhat large.  However, this system is left with $0.2~M_{\odot}$ of helium, which agrees with progenitor model predictions for SNe~Ic \citep{dessart+11,yoon+15}. If the SN~2017ein progenitor star evolved in a similar system with the luminosity and colors observed from PSF1, then it could still be relatively low mass, although the exact mass of the progenitor system and its overall abundances are still highly uncertainty.

From the light curve and spectra of SN~2017ein, we infer that this source is most similar to carbon-rich, low-luminosity SNe~Ic such as SN~2007gr as opposed to other SNe~Ic such as SN~2011bm.  This finding underscores the fact that the SN~2017ein progenitor star must have been hydrogen- and helium-deficient, but could also be relatively carbon-rich.  \citet{mazzali+10} and \citet{valenti+08} noted that the presence of strong carbon features at early times in SN~2007gr was consistent with an intrinsically high carbon-to-oxygen ratio in the progenitor star.

Our preferred maximum $V$-band absolute magnitude, assuming a host extinction of $A_{V}=1.2$~mag, suggests that SN~2017ein peaked at $M_{V}=-17.2\pm0.2$~mag.  This value is at the lower end of the luminosity function for SNe~Ic as a whole \citep{drout+11}, and is consistent with a relatively low mass of $^{56}$Ni ($<0.1~M_{\odot}$), also similar to SN~2007gr.  \citet{mazzali+10} point out that such a low $^{56}$Ni mass implies the explosion of a relatively low-mass CO core; in the case of SN~2007gr, this core likely resulted from a star with a main-sequence mass of $\sim15~M_{\odot}$ and a relatively low terminal mass \citep[see also][]{kim+15}.  \citet{yoon+10} suggest that these systems result from stars with a relatively low initial mass ($<25~M_{\odot}$) in order to explain the lack of helium and range of nickel masses.

This is in conflict with the $80$+$48~M_{\odot}$ binary model that provided the best match to the parameters of PSF1, where the $80~M_{\odot}$ star terminated at $15.4~M_{\odot}$.  It is possible that this is a result of systematic uncertainties in the models themselves.  Only 2 out of 12,664 BPASS models approached the properties of PSF1, whose extreme color and magnitude imply a massive, O- or B-type star.  Stars in this region of color-magnitude diagrams are not usually expected to explode, and so the lack of models that terminate here may reflect a physical limitation as much as systematic uncertainties in model parameters.  On the other hand, \citet{yoon+15} point out that in binary progenitor models for SNe~Ib/c, the terminal mass of the primary star is quite sensitive to the choice of metallicity, mass-loss, and mass-transfer prescriptions.  It is theoretically plausible that extreme mass transfer could produce a low terminal mass from a $80~M_{\odot}$ star \citep[see, e.g., Figure~10 in][]{yoon+15}.  However, such a scenario must be verified with late-time imaging to look for variations in the color and magnitude of the pre-explosion source.

\section{Conclusions}\label{sec:conclusions}

We present pre-explosion imaging and high-resolution imaging, photometry, and spectroscopy of the SN~Ic 2017ein.  We find:

\begin{enumerate}

	\item Spectra and light curves of SN~2017ein are remarkably similar to carbon-rich, low-luminosity SNe~Ic such as SN~2007gr and unlike SNe~Ic such as 2011bm.  At the same time, matching the continuum and peak $V$-band luminosity of SN~2017ein to SN~2007gr requires roughly $A_{V}=1.2$~mag of host extinction.  We also detect strong Na\I\ D absorption at the approximate redshift of NGC~3938.  These spectral characteristics suggest that the progenitor system contained very little hydrogen or helium, but also that it may have had an intrinsically high carbon abundance in its outer layers, as has been suggested for some WC stars.
	\item The location of SN~2017ein as determined from high-resolution laser guide star adaptive optics imaging is consistent with a single source in pre-explosion \hst/WFPC2 imaging.  The source is marginally extended in the \hst/WFPC2 images and there may be non-uniform background emission at this location.
	\item Accounting for the extended source and host extinction, photometry from the pre-explosion is consistent with single stars with masses up to $75~M_{\odot}$, but with a preferred mass of $55~M_{\odot}$.  However, most of these stars, which include O- and B-type supergiants and WN stars, are hydrogen-rich, and so are unlikely SN~Ic progenitor stars.
	\item Comparison to highly-stripped WR star binaries indicates that the only systems that match the colors and luminosity of PSF1 are WC+O and B star binaries.  We find that a $80$+$48~M_{\odot}$ BPASS model can explain some of the parameters of SN~2017ein and the pre-explosion counterpart and produces a star whose terminal state is roughly consistent with predictions of SN~Ic progenitor stars. Additional modeling is needed to explore the full ramifications of this evolutionary pathway and the precise terminal state of such a system.
	\item Nebular spectroscopy of SN~2017ein will be critical for measuring the true carbon abundance in the ejecta.  Late-time imaging of the site of SN~2017ein will also be important for measuring the extent to which the SN~2017ein progenitor star contributed to emission from the pre-explosion source.

\end{enumerate}

\bigskip\bigskip\bigskip
\noindent {\bf ACKNOWLEDGMENTS}
\smallskip
\footnotesize

We thank Raj Chowdhury and Bella Nguyen for help with Nickel observations.  We also thank David Coulter, C\'{e}sar Rojas-Bravo, and Matthew Siebert for help with Shane and Mayall observations.

The UCSC group is supported in part by NSF grant AST--1518052, the Gordon \& Betty Moore Foundation, the Heising-Simons Foundation, and by fellowships from the Alfred P.\ Sloan Foundation and the David and Lucile Packard Foundation to R.J.F.

Some of the data presented herein were obtained at the W. M. Keck Observatory, which is operated as a scientific partnership among the California Institute of Technology, the University of California, and NASA. The observatory was made possible by the generous financial support of the W. M. Keck Foundation. We wish to recognise and acknowledge the cultural significance that the summit of Mauna Kea has within the indigenous Hawaiian community. We are most fortunate to have the opportunity to conduct observations from this mountain.

Some of the data in this publication were calibrated using object catalogs from the Pan-STARRS1 Surveys.  The Pan-STARRS1 Surveys (PS1) and the PS1 public science archive have been made possible through contributions by the Institute for Astronomy, the University of Hawaii, the Pan-STARRS Project Office, the Max-Planck Society and its participating institutes, the Max Planck Institute for Astronomy, Heidelberg and the Max Planck Institute for Extraterrestrial Physics, Garching, The Johns Hopkins University, Durham University, the University of Edinburgh, the Queen's University Belfast, the Harvard-Smithsonian Center for Astrophysics, the Las Cumbres Observatory Global Telescope Network Incorporated, the National Central University of Taiwan, the Space Telescope Science Institute, the National Aeronautics and Space Administration under Grant No. NNX08AR22G issued through the Planetary Science Division of the NASA Science Mission Directorate, the National Science Foundation Grant No. AST-1238877, the University of Maryland, Eotvos Lorand University (ELTE), the Los Alamos National Laboratory, and the Gordon and Betty Moore Foundation.

The {\it Hubble Space Telescope} (\hst) is operated by NASA/ESA. The \hst\ data used in this manuscript come from programme GO-10877 (PI Li). Some of our analysis is based on data obtained from the \hst\ archive operated by STScI.

We acknowledge the use of public data from the {\it Swift} data archive.

Some of the data presented in this manuscript come from the Kitt Peak National Observatory (KPNO) 4-m telescope through programme 2017A-0306 (PI Foley).  KPNO is operated by the Association of Universities for Research in Astronomy, Inc. (AURA) under cooperative agreement with the National Science Foundation.

The Nickel and Shane telescopes are operated by the University of California and Lick Observatories.  Some of the data presented in this manuscript come from UCO/Lick programmes 2017Q2-N007, 2017Q3-N005 (PI Kilpatrick) and 2017A-S011, 2017B-S018 (PI Foley).

This work makes use of observations performed by the Las Cumbres Global Telescope Network through programme 2017AB-012 (PI Kilpatrick).

\textit{Facilities}: Keck (ESI/LRIS/OSIRIS), LCOGTN (SINISTRO), Mayall (KOSMOS), Nickel (Direct), Shane (Kast), {\it Swift} (UVOT)

\bibliography{2017ein}

\end{document}